\def\o{\omega}\def\r{\rho}\def\s{\sigma}\def\t{\tau}

\def\O{\Omega}

\def\de{\partial}
\def\inf{\infty}\def\mo{{-1}}\def\ha{{1\over 2}}

\def\arcth{{\rm arctanh}}

\def\ds{ds^2=}

\def\bh{black hole }

\def\bg{background }

\def\cc{coupling constant }

\def\ads{anti-de Sitter }
\def\RN{Reissner-Nordstr\"om }

\def\BR{Bertotti-Robinson }

\def\section#1{\bigskip\noindent{\bf#1}\smallskip}

\def\PL#1{Phys.\ Lett.\ {\bf#1}}
\def\PRL#1{Phys.\ Rev.\ Lett.\ {\bf#1}}
\def\CQG#1{Class.\ Quantum Grav.\ {\bf#1}}

\def\NC#1{Nuovo Cimento {\bf#1}}
 
\def\MPL#1{Mod.\ Phys.\ Lett.\ {\bf #1}}

\def\ref#1{\medskip\everypar={\hangindent 2\parindent}#1}
\def\beginref{\begingroup
\bigskip
\centerline{\bf References}
\nobreak\noindent}

\magnification=1200
{\nopagenumbers
\line{April 2001\hfil
}
\vskip40pt
\centerline{\bf Black holes and conformal mechanics}
\vskip40pt
\centerline{\bf S. Mignemi}
\vskip10pt
\centerline {Dipartimento di Matematica, Universit\`a di Cagliari}
\centerline{viale Merello 92, 09123 Cagliari, Italy}
\centerline{and INFN, Sezione di Cagliari}

\vskip40pt
\vskip120pt
{\noindent
We show how the motion of a charged particle near the horizon of an
extreme Reissner-Nordstr\"om black hole can lead to different forms
of conformal mechanics, depending on the choice of the time
coordinate.}
\vfil\eject}

Recently, it  has been shown that the motion of a charged particle near
the horizon of an extremal \RN \bh can be described (through dimensional
reduction to adS$_2$) by a model of
conformal mechanics [1], which, for great \bh mass reduces to the model
of De Alfaro, Fubini and Furlan (DFF) [2].
When quantized, the DFF model has a continuous spectrum of positive energy
eigenstates, but no normalizable ground state.

In [3,4], the absence of a ground state was interpreted as due to
a wrong choice of coordinates in the \bh metric, which do not
cover the entire manifold, and a different choice
was proposed, that solves the problem. The new time coordinate
corresponds to a different choice of conformal generators as
Hamiltonian for the conformal mechanics, and gives rise to a
'regularized' version of the DFF model, possessing a normalizable
ground state [2].

Although an algebraic proof of this fact was given in [3,4], it
was not shown how to derive the new Hamiltonian from the motion of
a charged particle in the near-horizon \RN background, and in
particular the relation between the parameters of the \bh and
those of the regularized DFF Hamiltonian remained obscure.

In this letter, we derive explicitly the regularized DFF
Hamiltonian from the motion of a charged particle in a suitably
parametrized adS$_2$ spacetime, in the 'non-relativistic' limit,
by taking into account higher order corrections in the inverse
mass of the black hole. We also discuss a further possible choice
of time coordinate, which however leads to a spectrum of energy
unbounded from below.

It is well known that the extreme \RN metric in the near-horizon limit,
$r/M\gg1$ can be put in the \BR form:
$$\ds-\left(r\over M\right)^2dt^2+\left(M\over r\right)^2dr^2+M^2d\O^2,
\eqno(1)$$
which is a direct product adS$_2\times S^2$. The motion of a test particle
in this background can be studied by considering the 2-dimensional \ads
section [1], namely
$$\ds-\left(r\over M\right)^2dt^2+\left(M\over r\right)^2dr^2.\eqno(2)$$
This provides a model of conformal mechanics in which the $SO(1,2)$ isometry
of the background spacetime is realized as a one-dimensional conformal
symmetry. The $so(1,2)$ algebra is generated by the Killing vectors $h=\de_t$,
$d=t\de_t-r\de_r$, $k=(t^2+M^4/r^2)\de_t-2tr\de_r$, which obey the commutation
relations,
$$[d,h]=-h,\qquad[d,k]=k,\qquad[h,k]=2d.\eqno(3)$$

Defining a new coordinate $q=-2M\sqrt{M\over r}$ the metric (2)
transforms into $$\ds-\left(2M\over q\right)^4dt^2+\left(2M\over
q\right)^2dq^2.\eqno(4)$$ The hamiltonian of a particle of mass
$m$ and charge $Q$ in this \bg is [1] $$H=\left(2M\over
q\right)^2\left(\sqrt{m^2+{q^2p_q^2\over4M^2}}-Q\right).
\eqno(5)$$
where $p_q$ is the momentum conjugate to $q$ and we have restricted
our attention to the radial motion. In the
'non-relativistic' limit [1], $M\to\inf$, $m-Q\to0$, with $(m-Q)M^2$ fixed,
the Hamiltonian (5) reduces to that of DFF quantum mechanics [2]:
$$H=\ha\left({p_q^2\over m}+{g\over q^2}\right),\eqno(6)$$
where the \cc $g$ is given by $8M^2(m-Q)$.

It is known that the DFF model has no ground state and its spectum is continuous.
This is due to the fact that the $so(1,2)$ generator $h$ associated to the
Hamiltonian (6) is noncompact, and can also be understood as a consequence of
the scaling invariance of the DFF model which does not allow a choice of a scale for
the energy.
From the \bh point of view, the absence of a ground state can be interpreted
as due to the existence of a fixed set for the Killing vector
$\de_t$, corresponding to its Killing horizon [3].
This problem can be remedied by adopting globally well defined coordinates,
$$u=\arctan{2rt\over r+r^\mo-rt^2},\qquad v=\ha(r-r^\mo+rt^2),$$
in terms of which the metric takes the form
$$\ds-\left[\left(v\over M\right)^2+1\right]du^2+\left[\left(v\over M\right)
^2+1\right]^{-1}dv^2.\eqno(7)$$
In this parametrization, the timelike Killing vector $\de_u$ corresponds to
the compact generator of $so(1,2)$, $\de_u=h+k$. It is known that this generator
admits a discrete spectrum with well-defined ground state [2].

In order to obtain the Hamiltonian for a charged particle which is conformal in
the appropriate limit, we make another change of coordinates,
which casts the metric in the form, inspired by (4),
$$\ds-A^2(x)du^2+A(x)dx^2.$$
This is obtained by defining
$$x=\int{dv\over\left[\left(v\over M\right)^2+1\right]^{3/4}}.$$
Unfortunately, this is an elliptic integral.
In order to obtain a closed form for the metric, we expand the integral
for $M/v\ll1$ (near-horizon limit), keeping track of the
first order corrections in $M/v$. It results
$$x\approx-2M\sqrt{M\over v}\left(1-{3M^2\over20v^2}\right),\quad{\rm i.e.}
\quad v\approx{4M^3\over x^2}\left[1-{3\over10}\left({x\over2M}\right)^4\right],$$
and then
$$ds^2\approx-\left({2M\over x}\right)^4\left[1+{2\over5}\left({x\over2M}
\right)^4\right]du^2+\left({2M\over x}\right)^2\left[1+{1\over5}\left(
{x\over2M}\right)^4\right]dx^2.\eqno(8)$$
The Hamiltonian for a charged particle moving in this background
is given by
$$H\approx\left(2M\over x\right)^2\left[1+{1\over5}\left({x\over2M}\right)^4
\right]\left[\sqrt{m^2+{x^2p_x^2\over4M^2}\left[1-{1\over5}\left(
{x\over2M}\right)^4\right]}-Q\right],\eqno(9)$$
which, in the 'non-relativistic' limit $M\to\infty$, $m-Q\to0$,
reduces to
$$H=\ha\left({p_x^2\over m}+{g\over x^2}+\o^2 x^2\right),\eqno(10)$$
where $g=8M^2(m-Q)$, $\o^2=(m-Q)/10M^2$.
This has the form of the regularized DFF Hamiltonian, which was introduced
in [2] in order to obtain a discrete energy spectrum with normalizable
ground state. The parameter $\o$ acts as an infrared cutoff, which breaks the
scaling invariance, setting the scale for the energy. In the approximation
of great $M$, $\o^2\ll g$.
In [3,4] was argued that (10) should be related to the motion of a charged
particle in the background (7), but no explicit derivation was given, and
in particular the value of $\o$ was left undetermined.
The spectrum of energy can now be obtained from the results of [2] and reads:
$$E_n=\sqrt{2m(m-Q)\over5}\left(2n+1+\sqrt{2{m-Q\over m}+{1\over4}}\right),$$
which is independent from $M$. The separation of levels tends to zero for
$m\to Q$.

To complete our discussion, we must consider another natural parametrization
of the two-dimensional anti-de Sitter space, which is obtained
through the change of coordinates [5]
$$\r=rt,\qquad\t=\arcth{r+r^\mo-rt^2\over r-r^\mo+rt^2}.$$
In these coordinates the metric takes the form
$$\ds-\left[\left(\r\over M\right)^2-1\right]d\t^2+\left[\left(\r\over M\right)
^2-1\right]^{-1}d\r^2,\eqno(11)$$
and the timelike Killing vector $\de_\t$ corresponds to the non-compact $so(1,2)$
generator $d$.
This parametrization has not been considered previously in this context.

As before, one can define a new coordinate
$$\s=\int{d\r\over\left[\left(\r\over M\right)^2-1\right]^{3/4}}
\approx-2M\sqrt{M\over\r}\left(1+{3M^2\over20\r^2}\right),$$
in terms of which the metric takes the form
$$ds^2\approx-\left({2M\over\s}\right)^4\left[1-{2\over5}\left({\s\over2M}
\right)^4\right]d\t^2+\left({2M\over\s}\right)^2\left[1-{1\over5}\left(
{\s\over2M}\right)^4\right]d\s^2.\eqno(12)$$
The Hamiltonian for a charged particle moving in this background
is given by
$$H\approx\left(2M\over\s\right)^2\left[1+{1\over5}\left({\s\over2M}\right)^4
\right]\left[\sqrt{m^2+{\s^2p_\s^2\over4M^2}\left[1-{1\over5}\left(
{\s\over2M}\right)^4\right]}-Q\right],\eqno(13)$$
which, in the 'non-relativistic' limit, reduces to
$$H=\ha\left({p_\s^2\over m}+{g\over\s^2}-\o^2\s^2\right),\eqno(14)$$
where $g$ and $\o$ are defined as before. We have again a regularized DFF model,
but now the harmonic potential has the wrong sign, leading again to the absence
of a ground state and to a spectrum unbounded from below [2]. This is in agreement
with the fact that the Hamiltonian
corresponds in this case to a non-compact generator $d$. From the \bh point of view,
this can again be related to the presence of a horizon at $\r=M$.

\beginref
\ref[1] P. Claus, M. Derix, R. Kallosh, J. Kumar, P.K. Townsend and A. Van Preoyen,
\PRL{81}, 4553 (1998);

\ref[2] V. De Alfaro, S. Fubini and G. Furlan, \NC{34A}, 569 (1976);

\ref[3] G. Gibbons and P. Townsend, \PL{B454}, 187 (1999);

\ref[4] R. Kallosh, {\tt hep-th 9902007}.

\ref[5] D. Christensen and R.B. Mann, \CQG{9}, 1769 (1992);
M. Cadoni and S. Mignemi, \MPL{A10}, 367 (1995).

\end